# Layer-specific connectivity revealed by diffusion-weighted functional MRI in the rat thalamocortical pathway

Daniel Nunes[1], Andrada Ianus[1,2], Noam Shemesh[1*]

[1]*Champalimaud Neuroscience Programme, Champalimaud Centre for the Unknown, Lisbon, Portugal*
[2]*Centre for Medical Image Computing, University College London, London, UK*

**\*Corresponding author:** Dr. Noam Shemesh, Champalimaud Neuroscience Programme, Champalimaud Centre for the Unknown, Av. Brasilia 1400-038, Lisbon, Portugal**.** E-mail: noam.shemesh@neuro.fchampalimaud.org . Phone number: +351 210 480 000 ext. #4467.

**Running title:** Coherent thalamocortical signals in diffusion-weighted fMRI

# Abstract

Investigating neural activity from a global brain perspective in-vivo has been in the domain of functional Magnetic Resonance Imaging (fMRI) over the past few decades. The intricate neurovascular couplings that govern fMRI's blood-oxygenation-level-dependent (BOLD) functional contrast are invaluable in mapping active brain regions, but they also entail significant limitations, such as non-specificity of the signal to active foci. Diffusion-weighted functional MRI (dfMRI) with relatively high diffusion-weighting strives to ameliorate this shortcoming as it offers functional contrasts more intimately linked with the underlying activity. Insofar, apart from somewhat smaller activation foci, dfMRI's contrasts have not been convincingly shown to offer significant advantages over BOLD, and its contrasts relied on significant modelling. Here, we study whether dfMRI could offer a better representation of neural activity in the thalamocortical pathway compared to its (spin-echo (SE)) BOLD counterpart. Using high-end forepaw stimulation experiments in the rat at 9.4 T, and with significant sensitivity enhancements due to the use of cryocoils, we show for the first time that dfMRI signals exhibit layer specificity, and, additionally, display signals in areas devoid of SE-BOLD responses. We find that dfMRI signals in the thalamocortical pathway *cohere* with each other, namely, dfMRI signals in the ventral posterolateral (VPL) thalamic nucleus cohere specifically with layers IV and V in the somatosensory cortex. These activity patterns are much better correlated (compared with SE-BOLD signals) with literature-based electrophysiological recordings in the cortex as well as thalamus. All these findings suggest that dfMRI signals better represent the underlying neural activity in the pathway. In turn, this may entail significant implications towards a much more specific and accurate mapping of neural activity in the global brain in-vivo.

# Introduction

The Blood Oxygenation Level Dependent (BOLD) mechanism has been at the heart of functional Magnetic Resonance Imaging (fMRI) since its inception[1-4] in the early 1990's. BOLD fMRI signals can act as surrogate markers for neural activity by harnessing intricate neurovascular couplings involving metabolic demand triggered by neural activity and ensuing fluctuations in blood oxygenation levels, as well as in hemodynamics[5-12]. BOLD fMRI has found diverse applicability in myriad disciplines due to its unique contrasts and noninvasive nature; indeed, BOLD-fMRI has been used, *inter alia*, to study neurovascular couplings in animals[13-15], to investigate functional connectivity[16] and cognition[17] in humans, and to explore functional deficits in disease and how they evolve with time[18].

BOLD fMRI's underpinnings are, however, somewhat removed from the underlying neuronal activity. That is, the neurovascular couplings are driven by neural activity, but they are governed by the vascular coupling. The spatial distribution of blood vessels[19,20], as well as the complex signaling pathways involved with activating the vasculature tree[21] will contribute to, and eventually determine, the shape of the BOLD response and hence fMRI's ability to faithfully represent the actual activity. In rat pups, for example, BOLD contrast is not observed until ~ day 18, when the neurovascular tree matures[22]; in physiologically unstable subjects, BOLD metrics may be unreliable[23]. In addition, BOLD's contrast – especially when imparted through gradient echo (GE) pulse sequences – is typically considered spatially unspecific due to recruitment of blood vessels far downstream from active foci[24-26], which often leads to overestimations of activated regions[27,28]. The application of spin-echo pulse sequences improves the specificity of the signals by filtering the larger veins and enhancing contrast from small microcapillaries that are more likely to be closer to the area of activity but comes at the expense of reduced sensitivity[29-31].

Clearly, the development of functional contrasts more intimately linked with the underlying neural activity would be a leap towards mapping brain function more directly and accurately. Many contrasts harnessing MRI's rich physics – that can give rise to many different types of contrast – have been suggested for this purpose[32]. Diffusion-weighted fMRI (dfMRI) is perhaps one of the most promising means suggested for overcoming some of BOLD-fMRI's limitations[33]. Diffusion MRI operates through the application of spatially-encoding gradients[34], separated by an observation time during which MRI-observable molecules can diffuse and encounter the microscopic boundaries in the tissue. The diffusion-weighted signal thus become imprinted with signatures of the underlying microstructure[35,36]. Already early on, diffusion weighted fMRI, with typically very low diffusion weighting, were used to provide insight into the BOLD mechanism itself and improve the spatial specificity of the signals[37-40]. However, given that a coupling between neural activity and mechanical properties has been long-since evidenced in intrinsic microscopy[41-46], and that the signal with higher diffusion weighting is typically associated with the extravascular space, Le Bihan and colleagues have postulated that diffusion MRI's sensitivity to microstructure could be used in a functional MRI setting and may reflect neural activity better than its BOLD counterpart[33,47,48]. Indeed, dfMRI contrasts exhibited a greater locality compared to BOLD-driven signals, both in humans[33,49] and in rodents[47,48], and functional changes were even observed in white matter upon stimulation[50]. However, dfMRI experiments have been also met with much criticism[51-54]: the more localized nature of activation was deemed an artifact of poorer signal to noise[53,54] and other studies questioned the very premise of the experiments, suggesting that the origins of the signal were not neural but BOLD-related[51,52,55]. To a great extent, the controversy over dfMRI signals can be pinpointed to: (1) low signal to noise levels, quite inherent to dMRI in general and dfMRI in particular; (2) the ensuing necessity of using elaborate statistical models[29,56] whose ground-truth or underpinnings are not necessarily known a-priori, which could impact the contrasts observed and affect signal interpretation; (3) perhaps most importantly, it was not shown, insofar, whether dfMRI signals actually provide "added-value" over BOLD-fMRI, in the context of mapping active networks in the brain.

Here, we endeavored to investigate whether dfMRI signals are more specific and whether they could map a known neural circuitry more closely than their SE-BOLD-fMRI counterparts. To overcome the sensitivity limitations[57], we harnessed a cryoprobe[58,59] at 9.4 T[60], thus boosting the sensitivity of the experiments to the point where pure data-driven analysis is possible, thereby realizing a fair comparison of functional signals. As a model system, we use the rat forepaw stimulation paradigm, where the circuitry is well-established from electrophysiology[13,61,62] and the ascending pathway is fully characterized (Figure 1A). In the specific case of mechanoreception in the rat forelimbs, somatosensory information is gated in the thalamic ventral posterolateral (VPL) nucleus, before reaching the cortex[63], which receives inputs mainly in layers IV[64,65] and V[66]. Thus, this model can be very useful to investigate whether dfMRI provides any added-value over BOLD-fMRI, mainly because BOLD-fMRI typically shows a non-local activation of the entire somatosensory cortex[67,68] and very rarely do signals specific to VPL emerge in BOLD-fMRI[9] (though unspecific thalamic signals are sometimes observed[69-71]). Thus, our hypothesis can be formulated as follows: if dfMRI more closely represents neural activity than BOLD-fMRI, its signals should be more specific to Layers IV/V, should show activation in VPL, and, those signals should *cohere* as they represent a chain of connected events. Our findings are all in line with our hypothesis, with dfMRI exhibiting preferential activation signals at Layer IV/V, and emerging at VPL (where SE-BOLD signals are not observed). Finally, strong coherence between VPL and Layer IV signals was demonstrated. Potential implications for future mapping brain activity, are discussed.

# Methods

All animal care procedures and experimental procedures were conducted in agreement with the European Directive 2010/63 and preapproved by the Champalimaud Animal Welfare Body.

### Animals

Long Evan rats, 8-10 weeks old (n=5), bred at the Champalimaud Vivarium, were housed in individually ventilated cages, in groups of two animals per cage, with food and water available *ad libitum*. The holding room was constantly monitored to maintain a controlled environment, at 23 ± 2°C, 12 h light–dark cycle.

### Animal preparation

In the day of the experiment, rats were induced into deep anesthesia with 5% isoflurane (Vetflurane, Virbac, France), and maintained under 2.5% isoflurane, while two stimulation electrodes (Dexter Electrostimulator 1.0, Hardware Platform, Champalimaud Research) were inserted into the left forepaw (between digits 1-2 and 4-5), whereupon animals were switched to medetomidine sedation[72] (Dormilan, Vetpharma Animal Health, Spain), 1 mg/ml solution diluted 1:10 in saline; bolus: 0.05 mg/kg, constant infusion: 0.1 mg/kg/h delivered via a perfusion pump (GenieTouch, Lucca Technologies, USA). The animal's temperature and respiration rate were continuously monitored using a rectal temperature probe and respiration sensor (SA Instruments Inc., USA), respectively, and $pCO_2$ was monitored using a transcutaneous monitoring system (TCM4 series, Radiometer, Denmark). In the end of the experiments, sedation was reverted by injecting the same amount of the initial bolus of 1:10 of atipamezole 1 mg/ml solution in saline (Antisedan, Vetpharma Animal Health, Spain).

### Stimulation paradigm

A stimulator built in-house was used to generate square waveforms for electrical stimulation at the left forepaw (Figure 1B). The stimulation protocol consisted of 45 seconds of rest, followed by 15 seconds stimulation with electrical pulses delivered to the left forepaw with a square waveform comprising 1.5 mA, 10 Hz and 3 ms stimulus duration. A total of 10 stimulation periods per experiment were used (Figure 1B).

### Functional Magnetic Resonance Imaging

A 9.4 T Bruker BioSpec scanner (Bruker, Karlsruhe, Germany) equipped with a gradient system producing up to 660 mT/m was used for all experiments. An 86 mm quadrature resonator was used for transmittance, while a 4-element array cryoprobe (Bruker, Fallanden, Switzerland) was used for signal reception [58,60]. Following localizer experiments, anatomical images were acquired using a RARE T2-weighted sequence (TE$_{eff}$/TE/TR=7.25/29/1800ms, RARE factor = 10, partial Fourier factor = 1.33, FOV = 16 x 16 mm, matrix size = 160 x 160, in-plane resolution = 100 μm x 100 μm, slice thickness = 0.75 mm). These images were used to place the four coronal slices of interest between +1.68 and -4.36 mm from Bregma.

For the all functional MRI experiments, a spin-echo echo planar imaging sequence (SE-EPI, Figure 1C) was used (TR/TE = 1500/45 ms, FOV = 16.1 x 16.1 mm, matrix size of 70 x 70, partial Fourier factor of 1.75, slice thickness = 1.5 mm, in-plane resolution 230 μm x 230 μm). To impart BOLD contrast (Figure 1C), the sequence was simply used as is, as it delivers T$_2$-weighted signals. To impart diffusion weighting, while avoiding the directionality of the standard diffusion acquisition, isotropic diffusion encoding (IDE)[73,74] gradient waveforms were implemented on the same exact sequence (Figure 1D), with b = 1500 s/mm$^2$, where $b = \int_0^t dt \left[ \gamma \int_0^t G(t')dt' \right]^2$, where *G(t)* represents the effective gradient waveform taking into account all refocusing pulses. A gradient separation (Δ) of 19.5 ms and a gradient duration (δ) of 14.2 ms were used (Figure 1D). The shape of the b-tensor was calculated using the magic angle (in Euler angles with respect to the gradient system, ζ = 54.3º, φ = 0º and θ = 0º, where ζ is the rotation axis (magic angle), while θ and φ reflect the polar and azimuthal angle in the gradient frame of reference) to generate spherical, i.e., isotropic diffusion encoding[74].

**Data analysis**

Both SE-BOLD and diffusion fMRI data were preprocessed using fMRat[75], a routine calling SPM12 in Matlab® (The Mathworks, Nattick, USA). Briefly, data were realigned, normalized and then slice-timing corrected prior to further data analysis. The realignment and normalization of data make use of interpolation algorithms implemented in SPM12, specifically a 4$^{th}$ degree B-spline, while slice-timing correction was performed using sinc interpolation.

*Region of interest analysis.* To avoid any assumptions on statistical models for data analysis, the preprocessed data were subject to region of interest (ROI) analysis. The ROIs were chosen based on known anatomy and drawn according to the Paxinos & Watson atlas[76], and corresponded to forelimb primary somatosensory cortex, FL S1 (slice 1, both in its entirety and, when indicated, only layer IV), and the ventral posterolateral area of the thalamus (VPL) (slice 4).

*Activation dynamics.* In each ROI, the detrended temporal evolution was plotted for single animals or the average over all animals, as indicated in the Figures. The averaged data plots mean±standard error of the mean (s.e.m.) of the data. The average cycle was calculated by averaging all stimulation epochs of all animals, so that the stimulation period was centered at each cycle.

*Signal distribution between epochs.* In ROI, the distribution of signals in active epochs vs. rest epochs was also plotted as a means of comparing the MRI signal in each period. Those distributions were then compared via a simple two-tailed Student's t-test with $p<0.05$ considered statistically significant. The cumulative distribution function was also extracted from these distributions for comparison.

*Pixel-by-pixel analysis.* We also aimed to generate activation maps from both SE-BOLD-fMRI and dfMRI, without relying on statistical models in the analysis. To this end, the preprocessed data were subject to both Fourier [77] and coherence analyses[78], as detailed below:

*Fourier analysis.* The paradigm is periodic, and therefore contains specific frequency elements and their harmonics. The paradigm was thus Fourier transformed, and the frequency elements identified in the magnitude spectrum. The SE-BOLD-fMRI and dfMRI signals were then also Fourier transformed, and the area under the frequencies corresponding to the first 2 components of the paradigm was computed. The ensuing activation maps simply reflect the area under those peaks.

*Coherence analysis.* To examine the relationship between active regions, coherence[78] analysis was performed. The temporal evolution of the ROIs drawn in VPL, FL S1, and Fl S1 layer IV regions were used as seeds to calculate the coherence magnitude between those seeds and every other pixel using Sun's method [78]. The integrals under the peaks in coherence spectra were computed.

*Mean diffusivity time course*: the raw temporal data $S_{b1500}$ and $S_{b0}$ extracted from an ROI placed in cortical layer 4 in each animal were prewhitened and low-pass filtered. For each animal, the stimulation epochs were averaged to produce the animal-specific averaged time course for each of the time courses at b=0 and b=1500 s/mm$^2$. For every animal, the mean diffusivity average time course was then calculated via $MD(t) = -\ln(S_{b1500}(t)/S_{b0}(t))/1500$. The animal-specific time-courses were then normalized, and the mean time course and its standard error were computed.

# Results

## Quality of raw data

The first objective of this study was to achieve sufficient sensitivity to be able to see the activation patterns with the naked eye even in a single animal and a single run. To assess the robustness of the experimental data in this study, Figures 1E and 1F show raw data from the SE-BOLD and dfMRI experiments, respectively, from a representative animal. The data were of high quality, without image artifacts, and, across the animals studied, the signal-to-noise ratio for SE-BOLD and dfMRI were 323±31, and 127±8, respectively (mean±standard deviation). The slight partial volume effects of the functional images can be judged when compared to the higher resolution anatomical images, which cover the same brain areas (Figure 1G).

## Raw temporal evolution of the signal in FL S1

Establishing whether dfMRI results could be observed across a single animal within a single stimulation epoch was the first goal of the study. Since the primary somatosensory cortex is the forepaw stimulation main target[65], Figure 2A shows FL S1 ROI (overlaid on the dfMRI image), specifically drawn in layer IV, which is known to receive the majority of inputs from such stimulation[64,65].

Figure 2B shows one representative dfMRI data traces arising from a representative animal. The signal changes on activation are clearly observed above the noise, even for this single trial, without averaging, filtering or otherwise data processing (other than detrending the data from the global drift). When the traces were averaged across only five animals (Figure 2C), the time-series profile becomes even cleaner and shows very strong signals that correspond to the paradigm. When those are summed to generate the average cycle (Figure 2D), a ~4% dfMRI signal increase is observed in the stimulation period. When the same ROI is placed over the SE-BOLD data (Figure 2E), the SE-BOLD signals show higher signal-to-noise contrast for both single animal (Figure 2F) and the average across animals (Figure 2G). The SE-BOLD average cycle signal (Figure 2H) is also cleaner than dfMRI, as expected, exhibiting less baseline noise. Nevertheless, both dfMRI and SE-BOLD fMRI show strong and robust functional signals in layer IV of somatosensory cortex. Note the similar amplitudes of the dfMRI and SE-BOLD fMRI activation signals.

To better quantify the differences between dfMRI and SE-BOLD signals, the distribution of signals in rest periods versus stimulation periods was evaluated (Figures 3A and 3B). The histograms clearly show an increased signal distribution at the stimulation period compared to the rest period for both dfMRI and SE-BOLD. The cumulative distribution functions shown in Figures 3C and 3D for dfMRI and SE-BOLD, respectively, separate rest and activity periods even more clearly. Both dfMRI and SE-BOLD signals differed with statistical significance between rest and stimulation conditions (Figure 3E, corrected p<0.0001 for both dfMRI and SE-BOLD).

**Raw temporal evolution of the signal in the VPL nucleus of the thalamus**

The second set of results investigated the occurrence of functional signals in the VPL nucleus in the thalamus, as the forepaw somatosensory pathway passes through this region. Data from an ROI drawn specifically in the VPL (Figure 4A) in a single animal (Figure 4B) are inconclusive, both for SE-BOLD and dfMRI. However, after averaging data from only five animals, clear signs of activity can be detected in dfMRI (Figure 4C) and the averaged cycle provides even more conclusive evidence towards signal increases in the VPL upon stimulation (Figure 4D). By contrast, when the (higher SNR) SE-BOLD experiments were performed, and the same ROI was used (Figure 4E), no signals could be distinguished from noise whether in single animals (Figure 4F), or in the signals summed from all five animals (Figure 4G). SE-BOLD's averaged cycle data also reveals no signs of signal differences in VPL between rest and stimulation conditions (Figure 4H).

To better quantify these effects, Figure 5 shows the histogram and cumulative distribution function analyses described above for the VPL ROI. The histogram distributions in dfMRI exhibit clear differences between rest and active periods (Figure 5A) while SE-BOLD histogram distributions reveal none (Figure 5B). The cumulative distribution function analysis (Figure 5C and 5D for dfMRI and SE-BOLD, respectively) reveals that only dfMRI cumulative distributions differ between rest and stimulation conditions (Figure 5C); no such difference was observed for the SE-BOLD CDFs (Figure 5D). Finally, the distribution means and standard deviations are plotted in Figure 5E. A statistically significant difference is observed in VPL only for the dfMRI experiment (corrected p <0.001) whereas the SE-BOLD experiment shows no statistically significant differences between rest and stimulation periods (corrected p=0.49).

**Mapping activity using spectral analysis**

Next, we turn to ROI and pixel-by-pixel activation mapping in the brain using a straightforward approach of Fourier analysis, which is made possible from the periodic nature of the paradigm (Figure 6A) and avoids fitting the data to specific assumed response functions[79,80]. The spectrum arising from the paradigm itself is presented in the lower panel of Figure 6A and contains the fundamental frequency (labeled 'F', ~ 0.016 Hz corresponding to one block per minute) as well as its harmonics. Figure 6B plots the frequency components obtained from the FL S1 ROI time-series for each method. Clearly, the first and second components are above the noise level and correspond to the same frequency components arising from the paradigm. Hence, only the fundamental frequency and second harmonic were considered for further analyses. Interestingly, in the VPL ROI, only dfMRI spectra contained signals in the fundamental frequency and the second harmonic, while SE-BOLD frequencies revealed no such components (Figure 6C).

To generate activation maps, the same Fourier analysis approach was simply performed voxel-by-voxel. Figure 6D shows the analysis for a single subject. For dfMRI, voxels corresponding to the input cortical layers IV and V were more correlated with the paradigm than the other cortical layers, while in SE-BOLD, this distinction was hardly possible to make, and a very large cortical area was observed as "active". Note that the thresholds used were identical for both methods, with the lower threshold assessed from the higher frequency components that contain only noise. When averaged across the five animals in this study, we obtained the final activation maps shown in Figure 6E: the dfMRI map shows a strong activation signal focused around laminae IV and V, with surrounding rims of lower activity; by contrast, the SE-BOLD-fMRI signals were much more uniform across the entire cortical region and were perhaps slightly more concentrated on the border between layers V and VI.

To evaluate whether thalamic activity could be mapped, the same procedure was applied to the slice containing VPL (Figure 6F). In the single-subject VPL analysis, VPL activity was very difficult to delineate in dfMRI (data not shown). However, when the data from the five animals were averaged, two thalamic nuclei corresponding to VPL and to the posterior medial nucleus (PoM, a thalamic nucleus involved in the adjustment of somatosensory cortical processing[81]), were identified as "active" (i.e., containing energy under the fundamental frequency and the second harmonic of the paradigm) in dfMRI. By contrast, no activation was observed in VPL for the (higher-SNR) SE-BOLD-fMRI experiments. Although multiple areas do appear "active" in these SE-BOLD maps, they are likely due to noise given that the signals in spectral domain were close to the noise levels in SE-BOLD (c.f. Figure 6C).

**Coherence in the rat thalamocortical pathway**

As described above, dfMRI exhibited activity in both VPL and FL S1, the areas most relevant to the forepaw stimulation paradigm. It is thus useful to evaluate whether these signals are functionally related to each other. To achieve this, we used coherence analysis[78], whose magnitude spectrum reflects a cross-correlation between a seed region and any other target. Figure 7A delineates the ROIs chosen for analyses. Figure 7B shows the coherence spectrum when the VPL ROI time course is used as the seed, and FL S1 time-series is used as the target. In the dfMRI experiments, a strong coherence was observed at the lower frequencies, as expected for functionally connected areas[78]. For the sake of completeness, Figure 7B (blue trace) also shows coherence plots arising from SE-BOLD, revealing, as expected, no significant coherence.

Finally, to investigate which brain regions cohere with the selected seeds, we evaluated coherence on a voxel-by-voxel level. Figure 7C shows that when the ROI drawn in FL S1 is used as the seed, the VPL is specifically highlighted in dfMRI coherence maps. Conversely, SE-BOLD signals originating in FL S1 did not cohere with thalamic nuclei (Figure 7D) or nearby thalamic nuclei. Similarly, when the VPL ROI was used as the seed, the strongest coherence in dfMRI was in the target area – the border of layers IV and V of FL S1 – along with signals cohering to a lesser extent in other layers in FL S1 (Figure 7E). As expected, SE-BOLD signals originating from VPL did not cohere with their downstream circuitry in the cortex (Figure 7F).

## Discussion

Neural activity is multifaceted, occurs on numerous scales and through myriad mechanisms, and therefore calls for the application of complementary methods, each providing information on different aspects of activity[78]. Mapping brain activity from a global perspective is a formidable challenge, but it provides a unique view of hierarchical brain structures and very different information compared to that obtained from, e.g., electrophysiological recordings[82] or fiber photometry[83], which observe a few cells at a time. BOLD-fMRI[1,4] has been the mainstay of such noninvasive global brain mappings, and it has indeed provided invaluable information on brain activity[5,7,8,21]. However, surmounting BOLD-fMRI's limits by mapping activity in a more specific way, could have a tremendous impact on how the brain can be studied in-vivo.

Diffusion-weighted fMRI has already been proposed by Le Bihan et al[33] as a means for mapping neural activity more directly than its BOLD counterpart. In fact, dfMRI was already considered in the 1990's as a means to investigate the origins of the BOLD response[38]. To understand these seemingly conflicting views of dfMRI (given that Le Bihan and co-workers assert that diffusion fMRI at high b-values is BOLD- independent[33,36,47,48]), it is necessary to consider the filter imposed on the signals. When rather weak diffusion weighting was used in the earlier studies, flow-related signals from relatively large vessels were filtered, thereby enabling an investigation of their impact on the BOLD response[38,39,84,85]. Indeed, application of weak diffusion weighting was shown to better localize BOLD effects[29,85-87]. At higher field strength the relative contribution of smaller vessels (on both intracellular and extracellular space) become more important[33,88-90] even with the native SE signal, and with strong diffusion weighting, the signals reflect a decrease in water diffusion within the tissue. Ample orthogonal evidence from experiments in slices[41,42] as well as in-vivo[44,45] using light scattering invokes the existence of a coupling between neural activity and mechanical properties in the tissue[91], on various spatial and temporal scales[91,92]; recently, further evidence to activity-related dynamic changes in tissue microstructure was given from super-resolution imaging[46]. Interestingly, neuronal swelling[93], astrocyte swelling[94], and swellings of axons[91] and boutons[95] have all been implicated with neural activity. However, given that it is difficult to assign diffusion MRI signals to specific cellular compartments/components, it is perhaps not surprising that much controversy still exists[51,53,54] on whether or not dfMRI signals actually reflect a microstructural (cell-swelling) effect or whether, though perhaps less likely[56], they arise from a "filtered" BOLD effect.

Our study aimed to SE-BOLD- and diffusion-fMRI in the most direct way possible, avoiding the application of complex statistical models that assume (unknown) response functions a-priori or deconvolution of data. This calls for high-quality data that would enable a data-driven analysis of both SE-BOLD and dfMRI signals. To achieve the required data quality, we took advantage of an arrayed cryogenic receive coil (cryoprobe) to boost signal-to-noise ratio (SNR) by approximately x2.5[58,60]. The basic principle underlying the SNR boost achieved with cryogenic coils relies mainly on lowering the electronic noise of the MR acquisition hardware. This fact has been shown theoretically[57] and demonstrated with NMR acquisitions using receiver coils and pre-amplifiers cooled to cryogenic temperatures[59]. Previous studies, using a similar cryoprobe engineered for mice, also reported an SNR increase of approximately x2.5, in comparison to a conventional room temperature receiver coil[60]. With this sensitivity, robust and reproducible cortical and thalamic signals were observed in dfMRI, which mapped the network with very high specificity.

Although our study does not address the origin of the dfMRI signals directly, it asks a similarly relevant question: *do dfMRI signals better represent our knowledge of the underlying neural activity compared with their BOLD counterparts?* Shih et al[62] characterized the electrophysiology of forepaw stimulation with parameters that are very close to those used in our paradigm. Using a current source density analysis, they found that most activity occurs on the border of layers IV and V, with weaker responses in layer VI and II-III. These electrical responses are in excellent agreement with our dfMRI findings, shown in Figure 6E, that revealed the strongest functional signals in layers IV and V, and diminished responses in layers VI and II-III; by contrast, the SE-BOLD signals observed in this study are much more homogeneously spread along the somatosensory cortex (Figure 6E), consistent with a much broader recruitment of vasculature in areas distant from the activation foci[24-26]. Interestingly, despite the lower signal-to-noise of dfMRI, the activation signal amplitudes are comparable to those of SE-BOLD-fMRI signals in the cortex. The more localized functional signals mentioned above in layers IV and V of FL S1 are also quite consistent with Tsurugizawa et al's dfMRI findings in rats[47]. However, these cortical signals should originate from the ascending stimulation sent by the thalamic VPL nucleus[64,65] (Figure 1A), which, to our knowledge, has not been yet investigated by dfMRI. Since the dfMRI signals evidenced activity in VPL (Figs 4 and 5), and since the areas are functionally connected[65], we hypothesized that they should cohere with each other – as expected from the neural activity in the somatosensory pathway[64,65]. Interestingly, dfMRI signals indeed exhibited very strong coherence between VPL and FL S1 Layers V and IV, and vice-versa, the entire somatosensory cortex cohered very well with VPL (Figure 7). This lends further credence to the notion that dfMRI signals (at high field) represent the network's neural activity quite faithfully. In fact, our study shows no SE-BOLD signals at all in VPL (c.f. Figure 6), and, as a consequence, also no coherence with S1. This observation cannot be attributed to signal-to-noise differences between SE-BOLD and dfMRI sequences, because dfMRI exhibited *lower* SNR, by a factor of almost 5, but it still evidenced signals not observed in the higher-SNR SE-BOLD-fMRI. The absence of VPL signals in SE-BOLD-fMRI is in line with, e.g., Keilholz et al[9], that observed BOLD-fMRI activity in thalamus only in <~10 % of rats studied, and, even then, the activity was not specific to VPL. On the other hand, several studies show BOLD activation of the thalamus in similar rat forepaw stimulation paradigms[70,71] using gradient-echo sequences. In GE, the sensitivity towards BOLD responses is higher, but the spatial specificity of the activated brain areas is much lower due to the sequence's tendency to highlight large draining veins and vessels downstream of the activated areas. Indeed, the above-mentioned reports[31,70,71] show activation of many thalamic nuclei that are not related with sensorial stimuli and the reproducibility of the VPL activity is shown to be quite poor[31]. One potential reason for the lack of clear VPL SE-BOLD signals, is its sparser, and differently organized, vasculature compared to the rodent cortex[96,97].

Rather than investigating the mechanism underlying dfMRI, this study was designed to provide high-quality data, which could be used nearly "as is" rather than be subjected to extensive modeling or filtering. The use of a cryoprobe, with approximately ×2.5 sensitivity enhancements[60], greatly contributed to the clarity of the data and its high SNR. Our ability to use a simple Fourier analysis to map the activation patterns obviated the need for statistical parametric mapping, which requires a-priori knowledge (or, more commonly, assumptions) on the response function (hemodynamic or diffusion). In addition, dfMRI signal distributions showed clear differences when simply plotted (Figures 3 and 5). It is also worth mentioning that the isotropic diffusion encoding scheme (IDE)[73,74] employed here benefits from removing potential directional dependence[51] (though, in auxiliary experiments we have not observed any significant orientational dependence, data not shown), as well as from mitigating cross-terms[73] with internal gradients due to their oscillatory nature, which in turn also reduces the sequence's potential sensitivity to BOLD effects when diffusion gradients are applied.

Despite that this study did not directly assess the mechanism underlying dfMRI signals, there are several putative explanations suggesting, at least, a shift of intra- to extra-vascular contrast mechanisms. For example, the emergence of diffusion-weighted functional signals in an area devoid of SE-BOLD-related functional signals, as observed in VPL (Figures 4 and 5) can be considered evidence against a vascular component contributing significantly to the dfMRI signals, as the contrast can only reflect diffusivity changes in the absence of BOLD responses at the non-diffusion-weighted signal. In other words, the diffusion-weighted signals, which have a lower signal-to-noise ratio than BOLD signals, are perhaps less likely to be a "filtered" BOLD response[51,53], because then one would expect to see the ("unfiltered") BOLD responses also in the much higher-SNR BOLD-fMRI experiments. In the cortex, it is perhaps more difficult to conclude whether the changes in dfMRI signals arise from changes in diffusivity or from internal gradients created by the susceptibility, as functional changes are observed both in b = 0 s/mm$^2$ (SE-fMRI) and b=1500 s/mm$^2$ data. The averaged time-course of the apparent mean diffusivity (MD) in S1 layer IV upon stimulation (Figure 8) reveals a clear MD decrease, locked to the neural activity but with a slightly different time course compared to SE-BOLD-fMRI and dfMRI, respectively. Still, it is worthwhile noting that, simplistically, for a simple 1-component system, $\ln\left(\frac{S_{IDE}(t)}{S_0(t)}\right) \propto -bD(t) + AG_i(t)G_{IDE}D(t)$ where $S_{IDE}$ and $S_0$ are diffusion weighted and non-diffusion weighted signals, respectively, $b$ is the b-value, $D$ is the diffusivity, $G_i$ is the internal gradient driven by susceptibility (e.g., reflects BOLD to some extent), and $A$ reflects sequence-specific timing constants. Dropping all inessential constant parameters, $\ln\left(\frac{S_{IDE}(t)}{S_0(t)}\right) \propto -D(t) + G_i(t)D(t)$. Therefore, the apparent MD time-dependence (Fig. 8) will only be identical to $D(t)$ if the internal gradient variation with time is zero (i.e. the cross-term cannot be necessarily neglected). The magnitude of the cross term will depend however on the overlap of internal gradient and IDE dephasing spectra, which will be quite small in our experiment, suggesting that $D(t)$ may be at least in part responsible for the MD changes in Figure 8. By contrast with the cortical data, $S_0(t)$ remained constant in the VPL, suggesting that $G_i(t)$ is small and the ensuing variations in mean diffusivity are likely due to modifications in $D(t)$. This result can indicate that dfMRI is in fact more sensitive to microstructural changes accompanying neuronal activity[90] or other features of the neuronal activity that are not reported by the SE-BOLD contrast. An alternative explanation for dfMRI signals relying entirely on hemodynamic signals, would involve a dynamic balance between negative and positive BOLD responses: it is not inconceivable that negative responses would be exactly balanced by positive responses in SE-BOLD fMRI, and, when strong diffusion weighting is applied, the negative contribution is filtered, giving rise to positive signals in VPL as observed in this study.

It is worth highlighting that this study does not tackle temporal aspects of dfMRI. Previous results from Tsurugizawa et al. in rodents[47], as well as others in humans[33,101], indicated that dfMRI signals typically peak faster than BOLD responses. In this study, we did not observe such temporal shifts since we aimed at mapping the spatial aspects of dfMRI, and thereby the temporal resolution was somewhat low. Thus, this study is not suitable for investigating such fast dynamics. Future studies with much higher temporal resolution are needed to investigate the temporal aspects of dfMRI and to provide insight into their correlation with underlying neural activity. These could be achieved using compressed sensing and/or sacrificing spatial resolution. Compared to other dfMRI studies[47,48] in rodents, this study used a relatively small number of animals (n=5); however, given the very high signal to noise of the experiment and the avoidance of using complicated statistical analyses, the study is sufficiently well powered, and the results were consistent along the different animals.

Finally, while the exact mechanism underlying dfMRI remains to be explored[29,40,52,55,98-100], our evidence suggests that whatever the mechanism, dfMRI functional signals are more specific to the circuitry, and therefore, may serve not only to highlight networks involved in task-based fMRI, but also perhaps more generally provide more genuine connectivity, e.g., in resting-state-dfMRI[48]. In addition, it should be noted that the data-driven analysis proposed here for the dfMRI signals should be applicable to human scans provided that the SNR is sufficiently high.

# Conclusions

Diffusion-weighted fMRI signals map neural activity more faithfully compared to their BOLD counterparts, at least from the perspective of known anatomical and functional connections. For the forepaw stimulation paradigm at 9.4 T, dfMRI signals are strongest in the border of layers IV and V, as expected from electrophysiology, and they cohere strongly with VPL signals, as expected from these anatomically and functionally connected regions, suggesting that the dfMRI signals are more intimately linked with the underlying activity than their hemodynamic-based SE-BOLD counterparts. Our data-driven analysis, which was able to reveal the time course of the apparent mean diffusivity, is independent of statistical modeling and thus avoids the inherent risk of bias in model selection. These findings are promising for future preclinical and clinical studies of neural activity and connectivity in the global brain.

# Acknowledgements

We would like to acknowledge the Vivarium of the Champalimaud foundation for help and support with the animals used in this study. We would like to thank Dr. Alfonso Renart for support with the coherence analysis and Dr. Cristina Chavarrias for assistance with the registration of the images. This study was supported by the European Research Council (ERC) under the European Union's Horizon 2020 research and innovation programme (Starting Grant, agreement No. 679058), and was also funded in part by EPSRC grant number M507970.

# Author contributions

N.S. designed the project and experiments. A.I. implemented the IDE sequence on the MRI scanner used in this study. D.N. performed experiments and the analysis. N.S. and D.N. wrote the manuscript.

# Financial interests

The authors declare no conflict of interest.

# References

1       Ogawa, S., Lee, T. M., Kay, A. R. & Tank, D. W. Brain magnetic resonance imaging with contrast dependent on blood oxygenation. *Proc Natl Acad Sci U S A* **87**, 9868-9872 (1990).
2       Ogawa, S. *et al.* Intrinsic signal changes accompanying sensory stimulation: functional brain mapping with magnetic resonance imaging. *Proc Natl Acad Sci U S A* **89**, 5951-5955 (1992).
3       Belliveau, J. W. *et al.* Functional mapping of the human visual cortex by magnetic resonance imaging. *Science* **254**, 716-719 (1991).
4       Kwong, K. K. *et al.* Dynamic magnetic resonance imaging of human brain activity during primary sensory stimulation. *Proc Natl Acad Sci U S A* **89**, 5675-5679 (1992).
5       van Zijl, P. C. *et al.* Quantitative assessment of blood flow, blood volume and blood oxygenation effects in functional magnetic resonance imaging. *Nat Med* **4**, 159-167 (1998).
6       Logothetis, N. K. What we can do and what we cannot do with fMRI. *Nature* **453**, 869-878, doi:10.1038/nature06976 (2008).
7       Attwell, D. & Iadecola, C. The neural basis of functional brain imaging signals. *Trends Neurosci* **25**, 621-625 (2002).
8       Logothetis, N. K. The underpinnings of the BOLD functional magnetic resonance imaging signal. *J Neurosci* **23**, 3963-3971 (2003).
9       Keilholz, S. D., Silva, A. C., Raman, M., Merkle, H. & Koretsky, A. P. Functional MRI of the rodent somatosensory pathway using multislice echo planar imaging. *Magn Reson Med* **52**, 89-99, doi:10.1002/mrm.20114 (2004).
10      Mukamel, R. *et al.* Coupling between neuronal firing, field potentials, and FMRI in human auditory cortex. *Science* **309**, 951-954, doi:10.1126/science.1110913 (2005).
11      Devor, A. *et al.* Coupling of the cortical hemodynamic response to cortical and thalamic neuronal activity. *Proc Natl Acad Sci U S A* **102**, 3822-3827, doi:10.1073/pnas.0407789102 (2005).
12      Goloshevsky, A. G., Silva, A. C., Dodd, S. J. & Koretsky, A. P. BOLD fMRI and somatosensory evoked potentials are well correlated over a broad range of frequency content of somatosensory stimulation of the rat forepaw. *Brain Res* **1195**, 67-76, doi:10.1016/j.brainres.2007.11.036 (2008).
13      Logothetis, N. K., Pauls, J., Augath, M., Trinath, T. & Oeltermann, A. Neurophysiological investigation of the basis of the fMRI signal. *Nature* **412**, 150-157, doi:10.1038/35084005 (2001).
14      Devor, A. *et al.* Coupling of total hemoglobin concentration, oxygenation, and neural activity in rat somatosensory cortex. *Neuron* **39**, 353-359 (2003).
15      Uhlirova, H. *et al.* Cell type specificity of neurovascular coupling in cerebral cortex. *Elife* **5**, doi:10.7554/eLife.14315 (2016).
16      Buchel, C., Coull, J. T. & Friston, K. J. The predictive value of changes in effective connectivity for human learning. *Science* **283**, 1538-1541 (1999).
17      He, B. J., Snyder, A. Z., Zempel, J. M., Smyth, M. D. & Raichle, M. E. Electrophysiological correlates of the brain's intrinsic large-scale functional architecture. *Proc Natl Acad Sci U S A* **105**, 16039-16044, doi:10.1073/pnas.0807010105 (2008).
18      Brier, M. R. *et al.* Loss of intranetwork and internetwork resting state functional connections with Alzheimer's disease progression. *J Neurosci* **32**, 8890-8899, doi:10.1523/JNEUROSCI.5698-11.2012 (2012).
19      Kennerley, A. J., Mayhew, J. E., Redgrave, P. & Berwick, J. Vascular Origins of BOLD and CBV fMRI Signals: Statistical Mapping and Histological Sections Compared. *Open Neuroimag J* **4**, 1-8, doi:10.2174/1874440001004010001 (2010).

20   Blinder, P. *et al.* The cortical angiome: an interconnected vascular network with noncolumnar patterns of blood flow. *Nat Neurosci* **16**, 889-897, doi:10.1038/nn.3426 (2013).

21   Attwell, D. *et al.* Glial and neuronal control of brain blood flow. *Nature* **468**, 232-243, doi:10.1038/nature09613 (2010).

22   Colonnese, M. T., Phillips, M. A., Constantine-Paton, M., Kaila, K. & Jasanoff, A. Development of hemodynamic responses and functional connectivity in rat somatosensory cortex. *Nat Neurosci* **11**, 72-79, doi:10.1038/nn2017 (2008).

23   Sicard, K. M. & Duong, T. Q. Effects of hypoxia, hyperoxia, and hypercapnia on baseline and stimulus-evoked BOLD, CBF, and CMRO2 in spontaneously breathing animals. *Neuroimage* **25**, 850-858, doi:10.1016/j.neuroimage.2004.12.010 (2005).

24   Turner, R. How much cortex can a vein drain? Downstream dilution of activation-related cerebral blood oxygenation changes. *Neuroimage* **16**, 1062-1067 (2002).

25   Diekhoff, S. *et al.* Functional localization in the human brain: Gradient-Echo, Spin-Echo, and arterial spin-labeling fMRI compared with neuronavigated TMS. *Hum Brain Mapp* **32**, 341-357, doi:10.1002/hbm.21024 (2011).

26   Harmer, J., Sanchez-Panchuelo, R. M., Bowtell, R. & Francis, S. T. Spatial location and strength of BOLD activation in high-spatial-resolution fMRI of the motor cortex: a comparison of spin echo and gradient echo fMRI at 7 T. *NMR Biomed* **25**, 717-725, doi:10.1002/nbm.1783 (2012).

27   Ugurbil, K., Toth, L. & Kim, D. S. How accurate is magnetic resonance imaging of brain function? *Trends Neurosci* **26**, 108-114, doi:10.1016/S0166-2236(02)00039-5 (2003).

28   Parkes, L. M. *et al.* Quantifying the spatial resolution of the gradient echo and spin echo BOLD response at 3 Tesla. *Magn Reson Med* **54**, 1465-1472, doi:10.1002/mrm.20712 (2005).

29   Lee, S. P., Silva, A. C., Ugurbil, K. & Kim, S. G. Diffusion-weighted spin-echo fMRI at 9.4 T: microvascular/tissue contribution to BOLD signal changes. *Magn Reson Med* **42**, 919-928 (1999).

30   Zhao, F., Wang, P. & Kim, S. G. Cortical depth-dependent gradient-echo and spin-echo BOLD fMRI at 9.4T. *Magn Reson Med* **51**, 518-524, doi:10.1002/mrm.10720 (2004).

31   Keilholz, S. D., Silva, A. C., Raman, M., Merkle, H. & Koretsky, A. P. BOLD and CBV-weighted functional magnetic resonance imaging of the rat somatosensory system. *Magn Reson Med* **55**, 316-324, doi:10.1002/mrm.20744 (2006).

32   Jasanoff, A. Bloodless FMRI. *Trends Neurosci* **30**, 603-610, doi:10.1016/j.tins.2007.08.002 (2007).

33   Le Bihan, D., Urayama, S., Aso, T., Hanakawa, T. & Fukuyama, H. Direct and fast detection of neuronal activation in the human brain with diffusion MRI. *Proc Natl Acad Sci U S A* **103**, 8263-8268, doi:10.1073/pnas.0600644103 (2006).

34   Stejskal, E. O. & Tanner, J. E. Spin Diffusion Measurements: Spin Echoes in the Presence of a Time-Dependent Field Gradient. *J Chem Phys* **42**, 288-+, doi:Doi 10.1063/1.1695690 (1965).

35   Le Bihan, D. Looking into the functional architecture of the brain with diffusion MRI. *Nat Rev Neurosci* **4**, 469-480, doi:10.1038/nrn1119 (2003).

36   Le Bihan, D. & Iima, M. Diffusion Magnetic Resonance Imaging: What Water Tells Us about Biological Tissues. *PLoS Biol* **13**, e1002203, doi:10.1371/journal.pbio.1002203 (2015).

37   Le Bihan, D., Turner, R. & MacFall, J. R. Effects of intravoxel incoherent motions (IVIM) in steady-state free precession (SSFP) imaging: application to molecular diffusion imaging. *Magn Reson Med* **10**, 324-337 (1989).

38   Silva, A. C., Williams, D. S. & Koretsky, A. P. Evidence for the exchange of arterial spin-labeled water with tissue water in rat brain from diffusion-sensitized measurements of perfusion. *Magn Reson Med* **38**, 232-237 (1997).

39  Lee, S. P., Silva, A. C. & Kim, S. G. Comparison of diffusion-weighted high-resolution CBF and spin-echo BOLD fMRI at 9.4 T. *Magn Reson Med* **47**, 736-741 (2002).
40  Jin, T., Zhao, F. & Kim, S. G. Sources of functional apparent diffusion coefficient changes investigated by diffusion-weighted spin-echo fMRI. *Magn Reson Med* **56**, 1283-1292, doi:10.1002/mrm.21074 (2006).
41  MacVicar, B. A. & Hochman, D. Imaging of synaptically evoked intrinsic optical signals in hippocampal slices. *J Neurosci* **11**, 1458-1469 (1991).
42  Andrew, R. D., Adams, J. R. & Polischuk, T. M. Imaging NMDA- and kainate-induced intrinsic optical signals from the hippocampal slice. *J Neurophysiol* **76**, 2707-2717, doi:10.1152/jn.1996.76.4.2707 (1996).
43  MacVicar, B. A., Feighan, D., Brown, A. & Ransom, B. Intrinsic optical signals in the rat optic nerve: role for K(+) uptake via NKCC1 and swelling of astrocytes. *Glia* **37**, 114-123 (2002).
44  Rector, D. M., Poe, G. R., Kristensen, M. P. & Harper, R. M. Light scattering changes follow evoked potentials from hippocampal Schaeffer collateral stimulation. *J Neurophysiol* **78**, 1707-1713, doi:10.1152/jn.1997.78.3.1707 (1997).
45  Rector, D. M., Rogers, R. F., Schwaber, J. S., Harper, R. M. & George, J. S. Scattered-light imaging in vivo tracks fast and slow processes of neurophysiological activation. *Neuroimage* **14**, 977-994, doi:10.1006/nimg.2001.0897 (2001).
46  Chereau, R., Saraceno, G. E., Angibaud, J., Cattaert, D. & Nagerl, U. V. Superresolution imaging reveals activity-dependent plasticity of axon morphology linked to changes in action potential conduction velocity. *Proc Natl Acad Sci U S A* **114**, 1401-1406, doi:10.1073/pnas.1607541114 (2017).
47  Tsurugizawa, T., Ciobanu, L. & Le Bihan, D. Water diffusion in brain cortex closely tracks underlying neuronal activity. *Proc Natl Acad Sci U S A* **110**, 11636-11641, doi:10.1073/pnas.1303178110 (2013).
48  Abe, Y., Tsurugizawa, T. & Le Bihan, D. Water diffusion closely reveals neural activity status in rat brain loci affected by anesthesia. *PLoS Biol* **15**, e2001494, doi:10.1371/journal.pbio.2001494 (2017).
49  Darquie, A., Poline, J. B., Poupon, C., Saint-Jalmes, H. & Le Bihan, D. Transient decrease in water diffusion observed in human occipital cortex during visual stimulation. *Proc Natl Acad Sci U S A* **98**, 9391-9395, doi:10.1073/pnas.151125698 (2001).
50  Spees, W. M., Lin, T. H. & Song, S. K. White-matter diffusion fMRI of mouse optic nerve. *Neuroimage* **65**, 209-215, doi:10.1016/j.neuroimage.2012.10.021 (2013).
51  Miller, K. L. *et al.* Evidence for a vascular contribution to diffusion FMRI at high b value. *Proc Natl Acad Sci U S A* **104**, 20967-20972, doi:10.1073/pnas.0707257105 (2007).
52  Kuroiwa, D. *et al.* Signal contributions to heavily diffusion-weighted functional magnetic resonance imaging investigated with multi-SE-EPI acquisitions. *Neuroimage* **98**, 258-265, doi:10.1016/j.neuroimage.2014.04.050 (2014).
53  Bai, R., Stewart, C. V., Plenz, D. & Basser, P. J. Assessing the sensitivity of diffusion MRI to detect neuronal activity directly. *Proc Natl Acad Sci U S A* **113**, E1728-1737, doi:10.1073/pnas.1519890113 (2016).
54  Williams, R. J., Reutens, D. C. & Hocking, J. Influence of BOLD Contributions to Diffusion fMRI Activation of the Visual Cortex. *Front Neurosci* **10**, 279, doi:10.3389/fnins.2016.00279 (2016).
55  Autio, J. A. *et al.* High b-value diffusion-weighted fMRI in a rat forepaw electrostimulation model at 7 T. *Neuroimage* **57**, 140-148, doi:10.1016/j.neuroimage.2011.04.006 (2011).
56  Aso, T. *et al.* An intrinsic diffusion response function for analyzing diffusion functional MRI time series. *Neuroimage* **47**, 1487-1495, doi:10.1016/j.neuroimage.2009.05.027 (2009).

57  Hoult, D. I. & Richards, R. E. The signal-to-noise ratio of the nuclear magnetic resonance experiment. 1976. *J Magn Reson* **213**, 329-343, doi:10.1016/j.jmr.2011.09.018 (2011).

58  Niendorf, T. *et al.* Advancing Cardiovascular, Neurovascular, and Renal Magnetic Resonance Imaging in Small Rodents Using Cryogenic Radiofrequency Coil Technology. *Front Pharmacol* **6**, 255, doi:10.3389/fphar.2015.00255 (2015).

59  Styles, P. *et al.* A high-resolution NMR probe in which the coil and preamplifier are cooled with liquid helium. 1984. *J Magn Reson* **213**, 347-354, doi:10.1016/j.jmr.2011.09.002 (2011).

60  Baltes, C., Radzwill, N., Bosshard, S., Marek, D. & Rudin, M. Micro MRI of the mouse brain using a novel 400 MHz cryogenic quadrature RF probe. *NMR Biomed* **22**, 834-842, doi:10.1002/nbm.1396 (2009).

61  Schouenborg, J., Kalliomaki, J., Gustavsson, P. & Rosen, I. Field potentials evoked in rat primary somatosensory cortex (SI) by impulses in cutaneous A beta- and C-fibres. *Brain Res* **397**, 86-92 (1986).

62  Shih, Y. Y. *et al.* Ultra high-resolution fMRI and electrophysiology of the rat primary somatosensory cortex. *Neuroimage* **73**, 113-120, doi:10.1016/j.neuroimage.2013.01.062 (2013).

63  Lopez-Bendito, G. & Molnar, Z. Thalamocortical development: how are we going to get there? *Nat Rev Neurosci* **4**, 276-289, doi:10.1038/nrn1075 (2003).

64  Petreanu, L., Mao, T., Sternson, S. M. & Svoboda, K. The subcellular organization of neocortical excitatory connections. *Nature* **457**, 1142-1145, doi:10.1038/nature07709 (2009).

65  Erzurumlu, R. S., Murakami, Y. & Rijli, F. M. Mapping the face in the somatosensory brainstem. *Nat Rev Neurosci* **11**, 252-263, doi:10.1038/nrn2804 (2010).

66  Constantinople, C. M. & Bruno, R. M. Deep cortical layers are activated directly by thalamus. *Science* **340**, 1591-1594, doi:10.1126/science.1236425 (2013).

67  Van Camp, N., Verhoye, M. & Van der Linden, A. Stimulation of the rat somatosensory cortex at different frequencies and pulse widths. *NMR Biomed* **19**, 10-17, doi:10.1002/nbm.986 (2006).

68  Weber, R., Ramos-Cabrer, P., Wiedermann, D., van Camp, N. & Hoehn, M. A fully noninvasive and robust experimental protocol for longitudinal fMRI studies in the rat. *Neuroimage* **29**, 1303-1310, doi:10.1016/j.neuroimage.2005.08.028 (2006).

69  Kim, Y. B., Kalthoff, D., Po, C., Wiedermann, D. & Hoehn, M. Connectivity of thalamo-cortical pathway in rat brain: combined diffusion spectrum imaging and functional MRI at 11.7 T. *NMR Biomed* **25**, 943-952, doi:10.1002/nbm.1815 (2012).

70  Lu, H. *et al.* Low- but Not High-Frequency LFP Correlates with Spontaneous BOLD Fluctuations in Rat Whisker Barrel Cortex. *Cereb Cortex* **26**, 683-694, doi:10.1093/cercor/bhu248 (2016).

71  Brynildsen, J. K. *et al.* Physiological characterization of a robust survival rodent fMRI method. *Magn Reson Imaging* **35**, 54-60, doi:10.1016/j.mri.2016.08.010 (2017).

72  Schlegel, F., Schroeter, A. & Rudin, M. The hemodynamic response to somatosensory stimulation in mice depends on the anesthetic used: Implications on analysis of mouse fMRI data. *Neuroimage* **116**, 40-49, doi:10.1016/j.neuroimage.2015.05.013 (2015).

73  Eriksson, S., Lasic, S. & Topgaard, D. Isotropic diffusion weighting in PGSE NMR by magic-angle spinning of the q-vector. *J Magn Reson* **226**, 13-18, doi:10.1016/j.jmr.2012.10.015 (2013).

74  Topgaard, D. Multidimensional diffusion MRI. *J Magn Reson* **275**, 98-113, doi:10.1016/j.jmr.2016.12.007 (2017).

75  Chavarrias, C. *et al.* fMRat: an extension of SPM for a fully automatic analysis of rodent brain functional magnetic resonance series. *Med Biol Eng Comput* **54**, 743-752, doi:10.1007/s11517-015-1365-9 (2016).

76  C.Watson, G. P. *The Rat Brain in stereotaxic coordinates*. 7th edition edn, (Academic Press, 2014).

77  Mitra, P. P. & Pesaran, B. Analysis of dynamic brain imaging data. *Biophys J* **76**, 691-708, doi:10.1016/S0006-3495(99)77236-X (1999).

78  Sun, F. T., Miller, L. M. & D'Esposito, M. Measuring interregional functional connectivity using coherence and partial coherence analyses of fMRI data. *Neuroimage* **21**, 647-658, doi:10.1016/j.neuroimage.2003.09.056 (2004).

79  Boynton, G. M., Engel, S. A., Glover, G. H. & Heeger, D. J. Linear systems analysis of functional magnetic resonance imaging in human V1. *J Neurosci* **16**, 4207-4221 (1996).

80  Glover, G. H. Deconvolution of impulse response in event-related BOLD fMRI. *Neuroimage* **9**, 416-429 (1999).

81  Wimmer, V. C., Bruno, R. M., de Kock, C. P., Kuner, T. & Sakmann, B. Dimensions of a projection column and architecture of VPM and POm axons in rat vibrissal cortex. *Cereb Cortex* **20**, 2265-2276, doi:10.1093/cercor/bhq068 (2010).

82  Brecht, M. *et al.* Novel approaches to monitor and manipulate single neurons in vivo. *J Neurosci* **24**, 9223-9227, doi:10.1523/JNEUROSCI.3344-04.2004 (2004).

83  Gunaydin, L. A. *et al.* Natural neural projection dynamics underlying social behavior. *Cell* **157**, 1535-1551, doi:10.1016/j.cell.2014.05.017 (2014).

84  Harshbarger, T. B. & Song, A. W. B factor dependence of the temporal characteristics of brain activation using dynamic apparent diffusion coefficient contrast. *Magn Reson Med* **52**, 1432-1437, doi:10.1002/mrm.20293 (2004).

85  Song, A. W. Diffusion modulation of the fMRI signal: early investigations on the origin of the BOLD signal. *Neuroimage* **62**, 949-952, doi:10.1016/j.neuroimage.2012.01.001 (2012).

86  Yablonskiy, D. A. & Haacke, E. M. Theory of NMR signal behavior in magnetically inhomogeneous tissues: the static dephasing regime. *Magn Reson Med* **32**, 749-763 (1994).

87  Song, A. W., Wong, E. C., Tan, S. G. & Hyde, J. S. Diffusion weighted fMRI at 1.5 T. *Magn Reson Med* **35**, 155-158 (1996).

88  Seehafer, J. U., Kalthoff, D., Farr, T. D., Wiedermann, D. & Hoehn, M. No increase of the blood oxygenation level-dependent functional magnetic resonance imaging signal with higher field strength: implications for brain activation studies. *J Neurosci* **30**, 5234-5241, doi:10.1523/JNEUROSCI.0844-10.2010 (2010).

89  Le Bihan, D. The 'wet mind': water and functional neuroimaging. *Phys Med Biol* **52**, R57-90, doi:10.1088/0031-9155/52/7/R02 (2007).

90  Le Bihan, D. Diffusion MRI: what water tells us about the brain. *EMBO Mol Med* **6**, 569-573, doi:10.1002/emmm.201404055 (2014).

91  Tasaki, I. & Byrne, P. M. Rapid structural changes in nerve fibers evoked by electric current pulses. *Biochem Biophys Res Commun* **188**, 559-564 (1992).

92  Tasaki, I. Rapid structural changes in nerve fibers and cells associated with their excitation processes. *Jpn J Physiol* **49**, 125-138 (1999).

93  Andrew, R. D. & MacVicar, B. A. Imaging cell volume changes and neuronal excitation in the hippocampal slice. *Neuroscience* **62**, 371-383 (1994).

94  Saly, V. & Andrew, R. D. CA3 neuron excitation and epileptiform discharge are sensitive to osmolality. *J Neurophysiol* **69**, 2200-2208, doi:10.1152/jn.1993.69.6.2200 (1993).

95  Sun, J. Y. & Wu, L. G. Fast kinetics of exocytosis revealed by simultaneous measurements of presynaptic capacitance and postsynaptic currents at a central synapse. *Neuron* **30**, 171-182 (2001).

96  Ghanavati, S., Yu, L. X., Lerch, J. P. & Sled, J. G. A perfusion procedure for imaging of the mouse cerebral vasculature by X-ray micro-CT. *J Neurosci Methods* **221**, 70-77, doi:10.1016/j.jneumeth.2013.09.002 (2014).
97  Errico, C. *et al.* Ultrafast ultrasound localization microscopy for deep super-resolution vascular imaging. *Nature* **527**, 499-502, doi:10.1038/nature16066 (2015).
98  Rudrapatna, U. S., van der Toorn, A., van Meer, M. P. & Dijkhuizen, R. M. Impact of hemodynamic effects on diffusion-weighted fMRI signals. *Neuroimage* **61**, 106-114, doi:10.1016/j.neuroimage.2012.02.050 (2012).
99  Duong, T. Q. *et al.* Microvascular BOLD contribution at 4 and 7 T in the human brain: gradient-echo and spin-echo fMRI with suppression of blood effects. *Magn Reson Med* **49**, 1019-1027, doi:10.1002/mrm.10472 (2003).
100 Nicolas, R., Gros-Dagnac, H., Aubry, F. & Celsis, P. Comparison of BOLD, diffusion-weighted fMRI and ADC-fMRI for stimulation of the primary visual system with a block paradigm. *Magn Reson Imaging* **39**, 123-131, doi:10.1016/j.mri.2017.01.022 (2017).
101 Aso, T., Urayama, S., Fukuyama, H. & Le Bihan, D. Comparison of diffusion-weighted fMRI and BOLD fMRI responses in a verbal working memory task. *Neuroimage* **67**, 25-32, doi:10.1016/j.neuroimage.2012.11.005 (2013).

# Figures

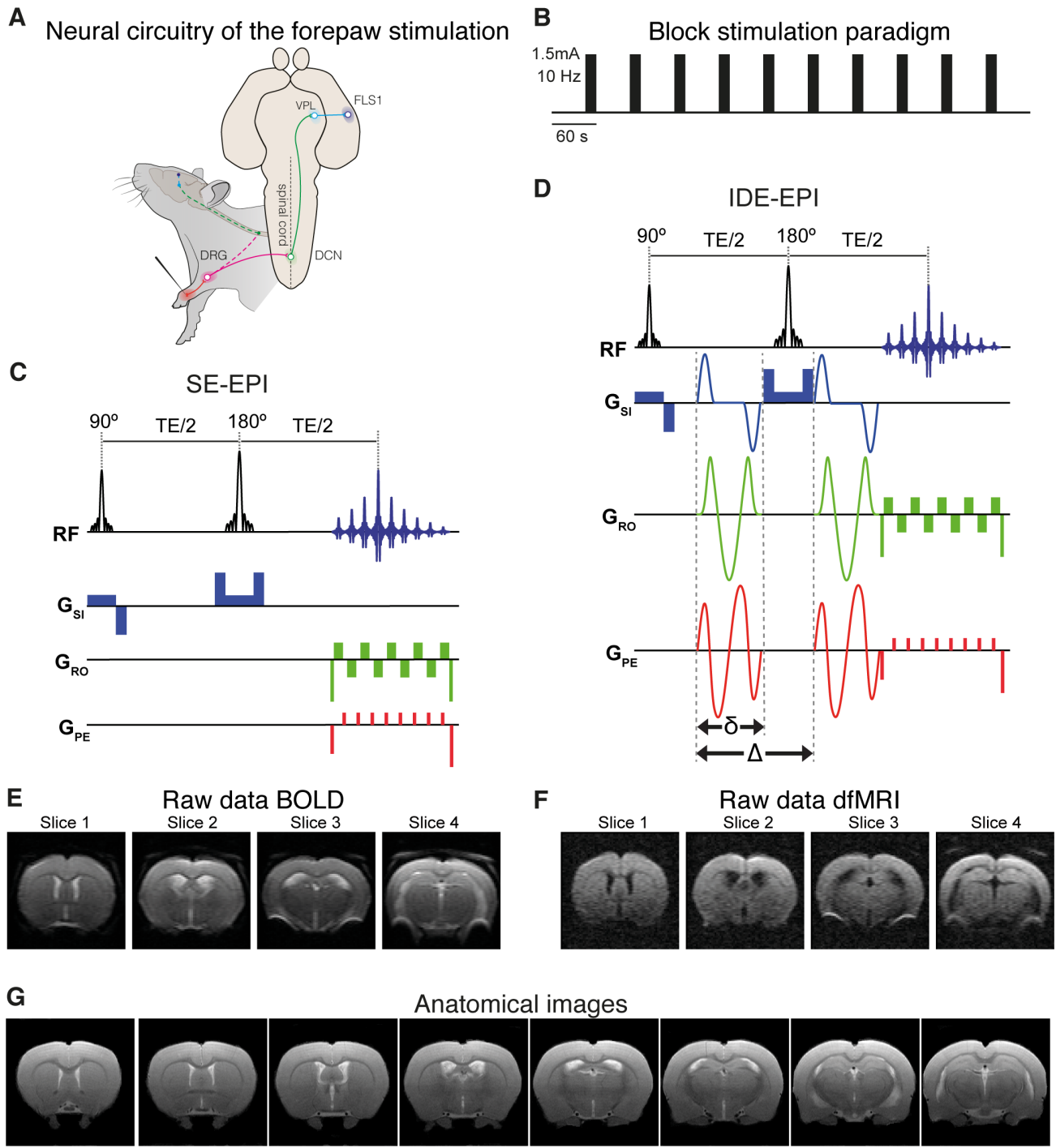

**Figure 1. Experimental design, pulse sequences, and raw data quality. (A)** Overview of the afferent neuronal network involved in the forepaw stimulation. An electrode pair was placed in the rat left forepaw. Upon electrical stimulation, sensory information is perceived by a specific group of afferent neurons (red) that project to the spinal cord. The information reaches the thalamus at the ventral posterolateral nucleus, through the dorsal column-medial lemniscus pathway (green). Axons project from the thalamic VPL to the forelimb primary somatosensory cortex (FL S1). **(B)** The stimulation paradigm consisted of 10 stimulation trains, 15 s each, interleaved by rest periods of 45 s. A rest period of 45 s was present at the beginning and end of the block stimulation paradigm. For stimulation, 1.5 mA pulses were used at a frequency of 10 Hz. **(C)** Spin-Echo Eco Planar Imaging (SE-EPI) sequence used for BOLD experiments. Note that it contains no diffusion gradients. **(D)** Isotropic Diffusion Encoding (IDE) EPI sequence used to acquire diffusion weighted images. The sequence is designed in such a way that the diffusion weighting is homogeneous in all directions. Δ and δ denote gradient separation and duration, respectively. **(E)** Representative example of raw images obtained in a SE-BOLD experiment. **(F)** Representative example of raw images obtained in an IDE-EPI experiment (b=1500 s/mm$^2$). **(G)** Representative anatomical images obtained using a RARE sequence. Abbreviations: DRG – dorsal root ganglion; DCN – dorsal central nucleus (dorsal column-medial lemniscus pathway); VPL – ventral posterolateral nucleus (of the thalamus); FL S1 – forelimb primary somatosensory cortex.

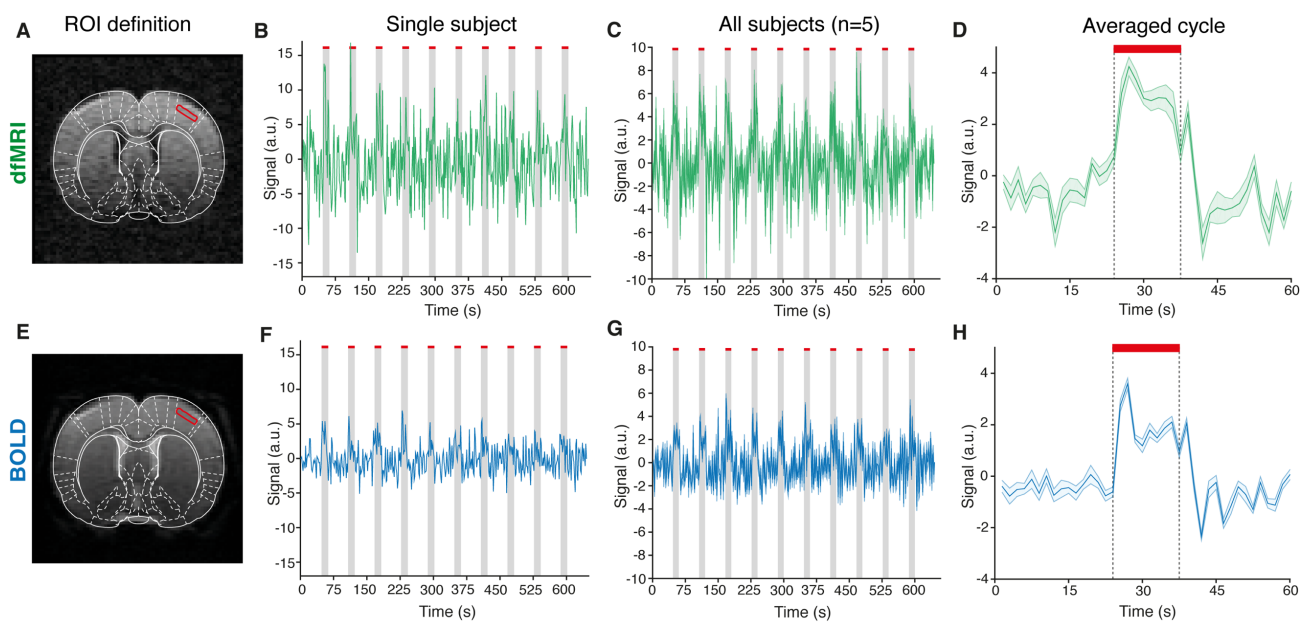

**Figure 2. Raw temporal evolution of the signal in the cortical layer IV of the FL S1. (A and E)** The ROI delineating the cortical layer IV of the FL S1 was designed based on the anatomical location (red color) for **(A)** dfMRI and **(E)** BOLD data. Representative traces of single **(B)** dfMRI and **(F)** SE-BOLD experiments. In single subjects the activity in layer IV of the FL S1 is evident for dfMRI and SE-BOLD experiments. For all animals tested (n=5), the average raw evolution time for **(C)** dfMRI and **(G)** SE-BOLD was calculated (mean±s.e.m.), showing clear signs of activity in dfMRI and SE-BOLD. (D and H) For all subjects (n=5), all stimulation epochs (10 per subject) were averaged using the ROIs corresponding to the anatomical region of cortical layer IV. The average cycle of **(D)** dfMRI and **(H)** SE-BOLD provides clear evidence for activity in cortical layer IV of the FL S1, which has similar temporal profiles. The red bars and grey columns denote stimulation epochs.

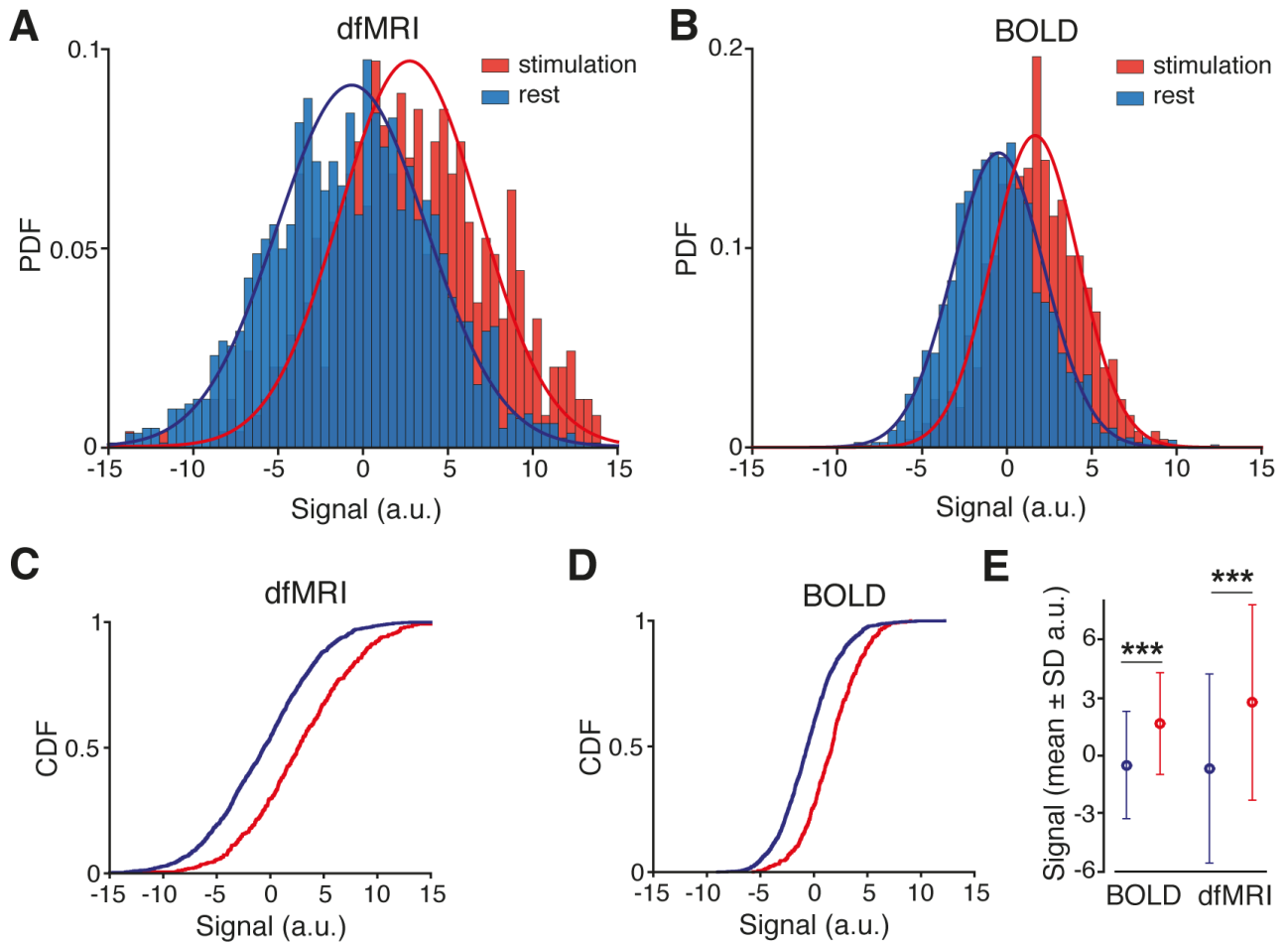

**Figure 3. Quantification of the activity in FL S1.** Histograms representing probability distribution function of the signals in rest periods (blue) versus stimulation periods (red) for **(A)** dfMRI and **(B)** BOLD signals and corresponding cumulative distribution functions in **(C)** dfMRI and **(D)** BOLD. This data comprises the rest and stimulation periods of all subjects (n=5; rest periods comprises a total of 1650 data points and stimulation periods comprises a total of 500 points). This analysis shows that the rest periods differentiate well from the stimulation periods. **(E)** Both dfMRI and BOLD signals differed with statistical significance between rest and stimulation conditions (mean±S.D.; two-samples t-test; t=-13.58, p<0.0001).

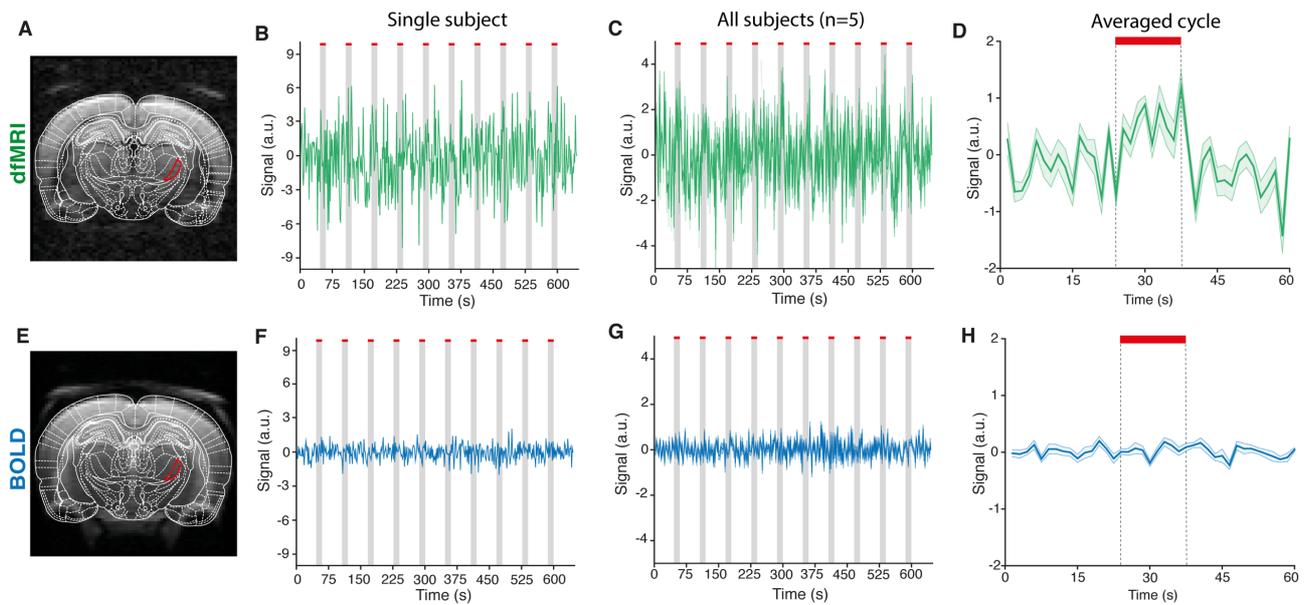

**Figure 4. Raw temporal evolution of the signal in VPL (thalamus). (A and E)** The ROI delineating the VPL was designed based on the anatomical location (red color) for **(A)** dfMRI and **(E)** SE-BOLD data. Representative traces of single **(B)** dfMRI and **(F)** SE-BOLD experiments. For all animals tested (n=5), the average raw evolution time for **(C)** dfMRI and **(G)** SE-BOLD was calculated (mean±s.e.m.), showing clear signs of activity in dfMRI. (D and H) For all subjects (n=5), all stimulation epochs (10 per subject) were averaged using the ROIs corresponding to the anatomical region of VPL. The average cycle of **(D)** dfMRI provides evidence for activity in VPL, while **(H)** in SE-BOLD that activity is not detected. The red bars and grey columns denote stimulation epochs.

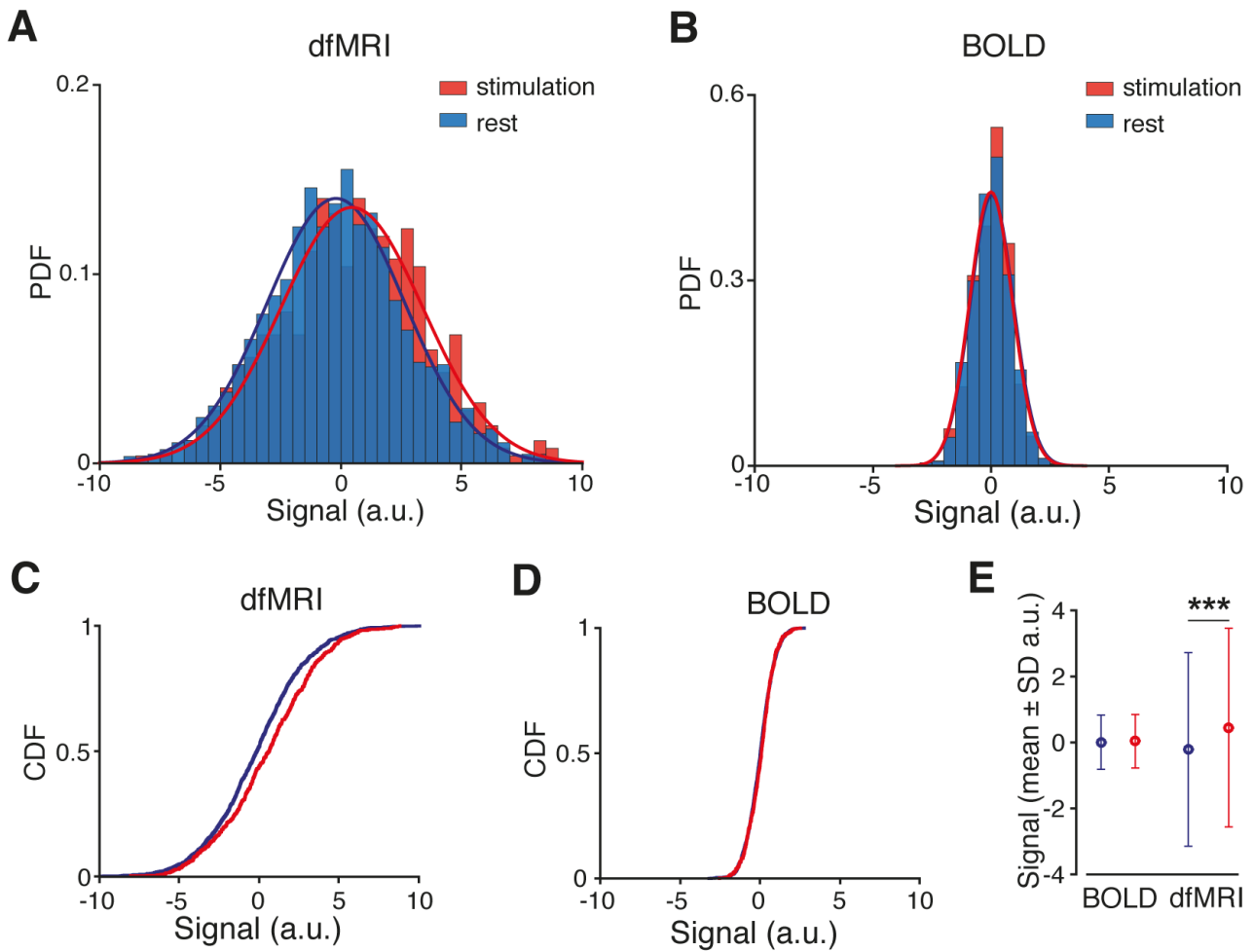

**Figure 5. Quantification of the activity in VPL (thalamus).** Histograms representing probability distribution function of the signals in rest periods (blue) versus stimulation periods (red) for **(A)** dfMRI and **(B)** BOLD signals and corresponding cumulative distribution functions in **(C)** dfMRI and **(D)** BOLD. This data comprises the rest and stimulation periods of all subjects (n=5; rest periods comprises a total of 1650 data points and stimulation periods comprises a total of 500 points). This analysis shows that the rest periods differentiate well from the stimulation periods in dfMRI but not in BOLD data. **(E)** The dfMRI signals differed with statistical significance between rest and stimulation conditions (mean±S.D.; two-samples t-test; t=-4.38, p<0.0001), while BOLD signals did not (mean±S.D.; two-samples t-test; t=-0.68, p=0.49).

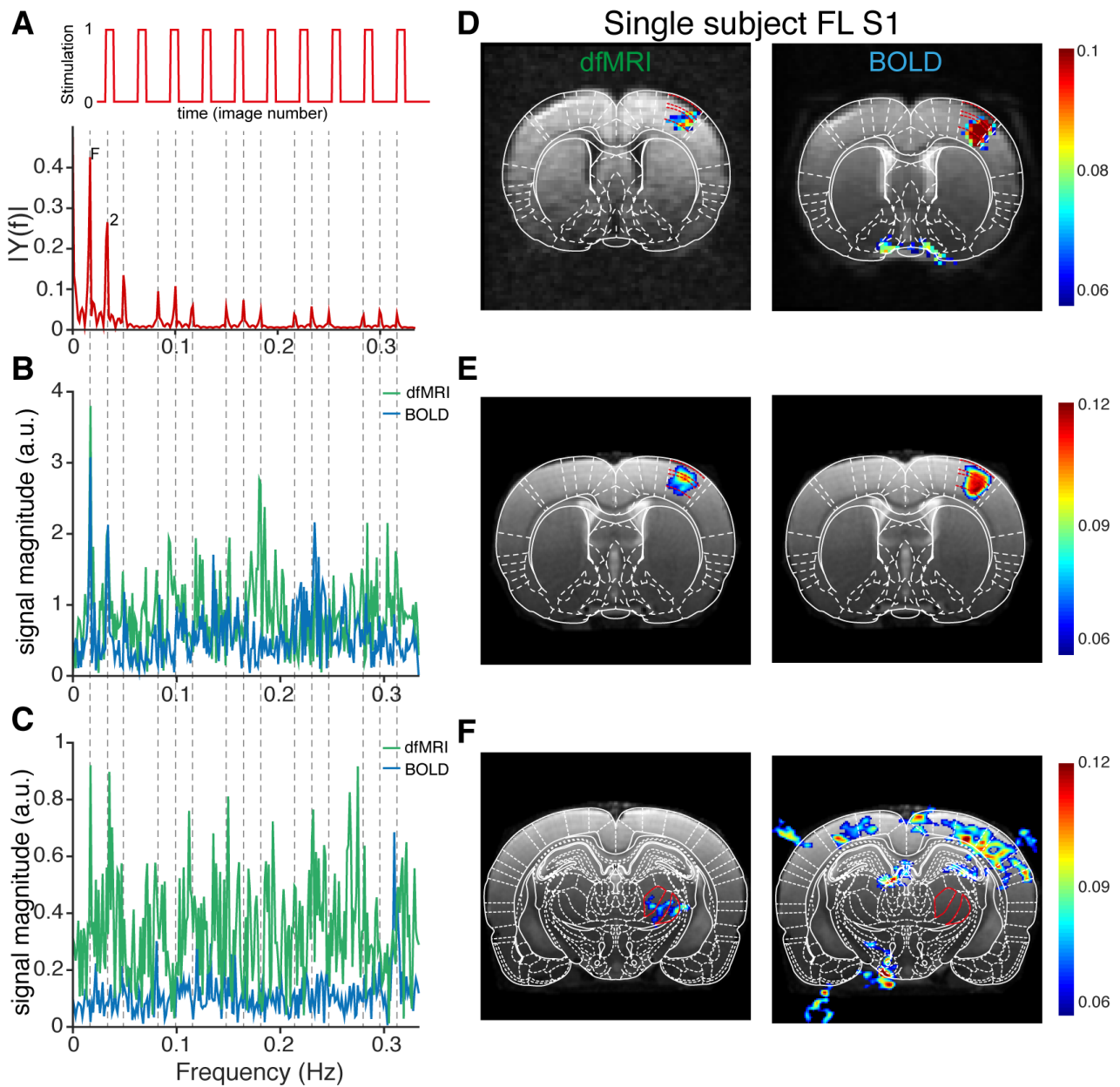

**Figure 6. Fourier analysis. (A)** Stimulation paradigm and corresponding Fourier spectrum, composed of a fundamental frequency at 0.016 Hz and its harmonics. (B and C) Frequency spectra of temporal evolutions in **(B)** FL S1 and **(C)** VPL nucleus of the thalamus. Only the fundamental frequency and the second harmonic separate well from noise. (D-F) Spectral voxel-by-voxel analysis maps the brain activity in S1, **(D)** in a single representative subject and **(E)** in the average of all subjects. **(F)** Average maps to show thalamic activity. In dfMRI it is possible to observe thalamic activity in VPL, corresponding to anatomical defined thalamic nuclei involved projecting somatosensory information to S1, and in PoM that is involved in the control of the somatosensory cortical processing. In the SE-BOLD images these brain regions are not revealed. Color maps represent spectral amplitudes.

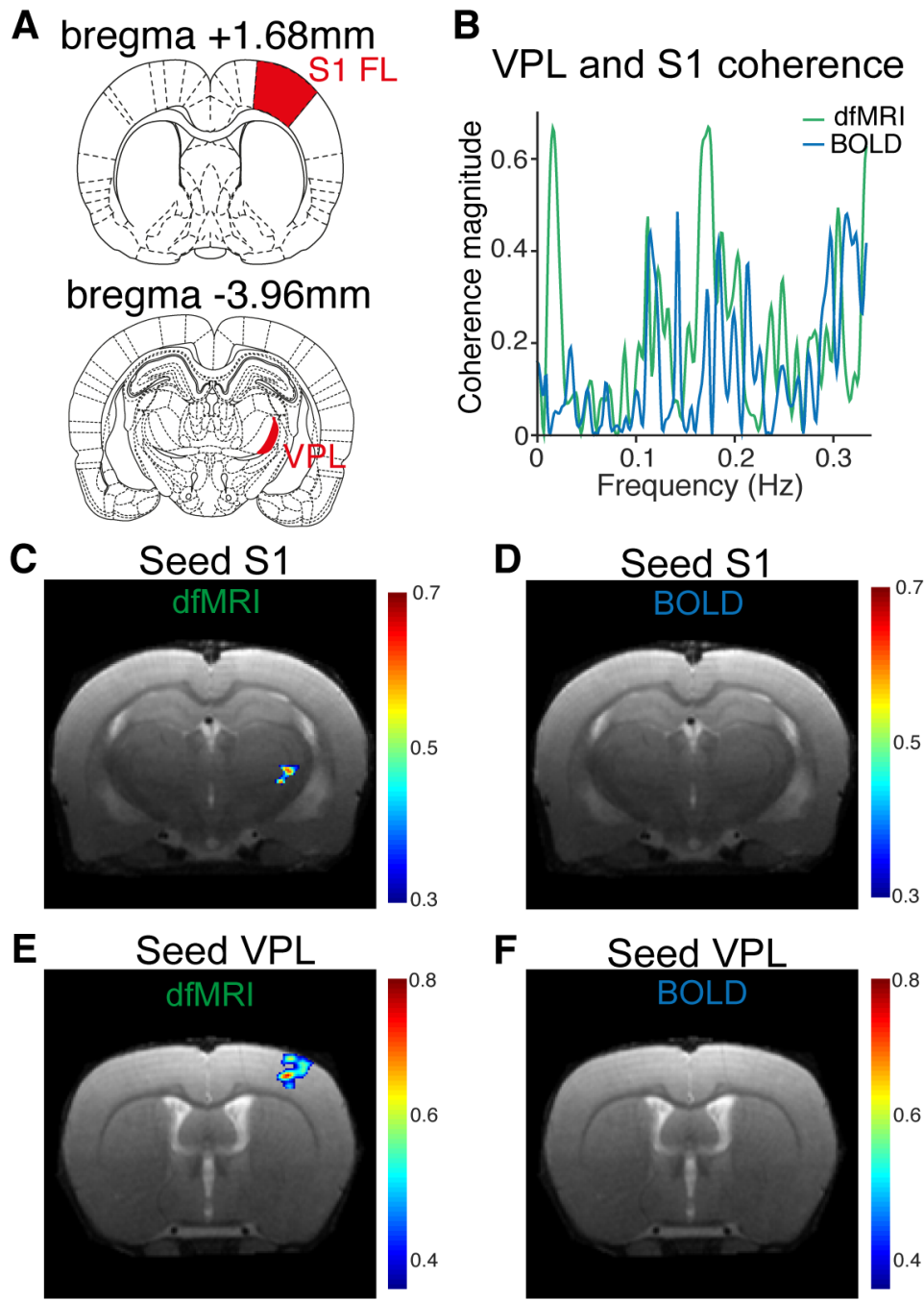

**Figure 7. Coherence analysis. (A)** The ROIs delineating VPL nucleus of the thalamus and the cortical layer 4 of the FL S1, based on the anatomy, were used to calculate the coherence between them. Note that these regions lay approximately 5mm apart. **(B)** Coherence graph representing the coherence between VPL and FL S1 for the dfMRI signal (green) and BOLD signal (blue) at the stimulation frequency (0.0166Hz). (C-D) Using the entire FL S1 cortical region as seed to map thalamic activity, **(C)** it is possible to observe that the VPL nucleus of the thalamus is highly correlated with the activity in FL S1 in dfMRI data, **(D)** but not in BOLD data. (E-F) Using the VPL nucleus of the thalamus as seed to map cortical activity, **(E)** it is possible to observe that the cortical layers of the FL S1 is highly correlated with the activity in VPL in dfMRI data, **(F)** but not in BOLD data. The color scale activity maps represent coherence magnitude overlaid on anatomical images.

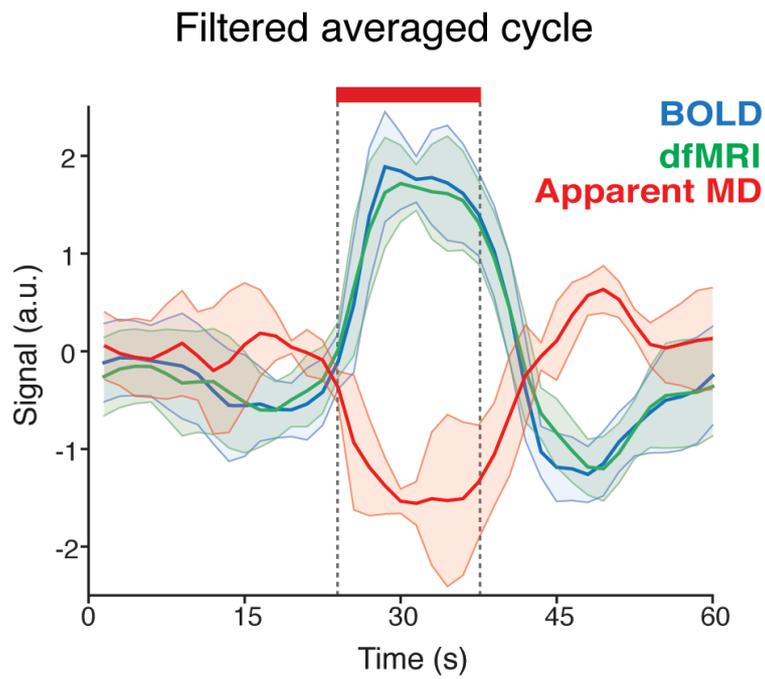

**Figure 8. Mean diffusivity changes in cortical layer 4 due to stimulation.** The raw temporal evolution of the signal in cortical layer 4 of the FL S1 region was used to calculate mean diffusivity changes due to stimulation. All subjects (n=5) and stimulation epochs (n=10 per subject) were averaged to calculate the filtered average cycle. Color code: blue – filtered averaged BOLD data, green – filtered averaged diffusion-weighted data (dfMRI), red – apparent mean diffusivity (MD).